\documentstyle[preprint,aps]{revtex}
\input psfig.sty

\tightenlines

\newcommand\be{\begin{equation}}
\newcommand\ee{\end{equation}}
\newcommand\bea{\begin{eqnarray}}
\newcommand\eea{\end{eqnarray}}
\newcommand\bean{\begin{eqnarray*}}
\newcommand\eean{\end{eqnarray*}}

\newcommand\ds\displaystyle

\newcommand\vaps{\varepsilon}
\newcommand\si{\sigma}
\newcommand\la{\lambda}
\newcommand\de{\delta}

\begin{document}

\draft

\title{A new approach to electromagnetic wave tails on a curved
spacetime}

\author{
  Romi Mankin}%
 \address{Department of Theoretical Physics, Tallinn Pedagogical 
  University, 25 Narva Road, 10120 Tallinn, Estonia}
\author{
  T\~onu Laas%
  \footnote{Electronic address: tony@tpu.ee}}
 \address{Department of Theoretical Physics, Tallinn Pedagogical 
  University, 25 Narva Road, 10120 Tallinn, Estonia\\
  and Institute of Theoretical ~Physics, Tartu University, 4 T\"ahe 
  Street, 51010 Tartu, Estonia}
\author{
  Risto Tammelo%
  \footnote{Electronic address: tammelo@physic.ut.ee}}
\address{Institute of Theoretical ~Physics, Tartu University, 4
T\"ahe 
  Street, 51010 Tartu, Estonia}

\date{\today}

\maketitle

\begin{abstract}
We present an alternative method for constructing the exact and approximate
solutions
of electromagnetic wave equations whose source terms are arbitrary
order multipoles on a curved spacetime.
The developed method is based on the higher-order Green's functions
for wave equations which are defined as distributions that satisfy
wave equations with the corresponding order covariant derivatives of
the Dirac delta function as the source terms.
The constructed solution is applied to the study of various geometric
effects on the generation and propagation of electromagnetic wave
tails to first order in the Riemann tensor.
Generally the received radiation tail occurs after
a time delay which represents geometrical backscattering by the central 
gravitational source.
It is shown that for an arbitrary weak gravitational field it is valid that
the truly nonlocal wave-propagation 
correction (the tail term) has a universal form which is independent
of multipole structure of the gravitational source.
In a particular case when the electromagnetic radiation pulse is generated by
the
wave source during a finite time interval, the structure of the wave tail at
the time
after the direct pulse has passed the
gravitational source is in the first approximation independent of the higher
multipole
moments of the source of gravitation, including the angular
momentum.
These results are then applied to a compact binary system. It follows that 
under certain conditions the tail energy can be a noticeable fraction 
of the primary pulse energy, namely, it is shown that for a
particular model the energy carried away by the tail can amount to
$10 \%$ of
the energy of the low-frequency modes of the direct pulse. 
The present results indicate that 
the wave tails should be 
carefully considered in energy calculations of such systems
and that the delay effect of the
wave tails may be of great importance for their observational
detection.

\end{abstract}

\section{Introduction}

The problem of radiation propagation on a curved spacetime has
been studied since the birth of general relativity, both
as a matter of principle and as a basis of observational
predictions. A field satisfying a hyperbolic
differential equation (wave equation) propagates not only on the
characteristic surfaces (light-cones), but also inside the
light-cone in the form of wave tails which
violate the Huygens' principle \cite{hadamard,gynther}.
Physically speaking, the wave tails arise because the radiation is
backscattered by the spacetime curvature. In certain cases
backscattering can influence observations as it weakens and disperses
sharp initial pulses. 
For instance, the electromagnetic (or gravitational) radiation from a
pulsed source in the vicinity of a massive body reaching an
asymptotic observer is received as two distinct pulses: one arriving
along the direct route and the other, the tail, effectively being
scattered off by the central body \cite{roe,peters1,peters2}. 

Recently the wave tails have become to be recognized as factors in the
planned observational detection of the gravitational waves by
forthcoming laser interferometric detectors 
\cite{blanchet1,blanchet2,blanchet3,bonnor,leonard,blanchet4}. 
It has also been shown that the wave tails play an important role in
the generation of gravitational waves by the orbital inspiral of a
compact binary system \cite{blanchet5,poisson,cutler}.
The close relationship between the generation of the wave tails
and gravitational focusing has been demonstrated in Ref.
\cite{thorne}.

The process of wave propagation on a curved spacetime being quite 
complicated, usually the wave equation is solved in the weak-field 
and slow-motion limit by the method of successive approximations, 
using  multipole expansion in one way or another (see e.g.
\cite{bonnor2,bardeen,couch,hunter}).
A general solution of the electromagnetic wave equation in a curved
space was constructed by DeWitt and Brehme \cite{dewitt} in terms 
of the Green's function, using the Hadamard procedure
\cite{hadamard}. The general solution demonstrates scattering of the 
electromagnetic radiation by the spacetime curvature.
An alternative technique for investigating electromagnetic radiation
in the Schwarzschild and Kerr metrics was worked out
following the Regge-Wheeler approach to metric 
perturbations \cite{regge,price,leaver}. This technique relies on
an expansion into generalized spherical harmonics and is 
especially useful when radiation emanates from a given multipole
moment of the source. Herein, for a point charge an infinite sum
emerges. The benefit, however, is that the obtained solution is also
valid for the strong-field region. 

We have recently developed a new method 
\cite{mankin3,mankin4,mankin8}
for calculating the exact solutions of scalar and tensor wave
equations whose source terms are arbitrary-order multipoles on a curved 
spacetime. The developed method is based on the higher-order fundamental
solutions (Green's functions) for wave equations which are defined in 
our paper \cite{laas} as the distributions that satisfy wave equations
with the corresponding order covariant derivatives of the Dirac delta 
function as the source terms.
Provided that the classical Green's function
and the multipole expansion of the source term are given,
there is no need for a small expansion parameter within the framework of
our formalism and in certain cases we can even find the exact multipole
solutions for  a strong field. Our approach also enables us to find an
approximate solution provided that an approximate form of the classical Green's 
function is known which is the case for most spacetimes. Moreover, it proves
to be more advantageous to apply our algorithm \cite{mankin4,mankin8},
instead of the traditional approach of successive approximations,
as the amount of computations involved is considerably reduced 
and, as will be demonstrated below, some features unrevealed by
the successive approximation methods are brought forward.
It is worthwhile to point out that, as distinct from most of papers
dedicated to the topics of wave tails in which the source of the 
gravitational field is regarded as a point mass, within the 
framework of our approach \cite{mankin3,mankin4} the extension
of the source of gravitation is finite. The last circumstance enables 
us to avoid in computations the additional non-physical singularity, the
regularizing of which may bring on difficulties in interpreting the results.

On the basis of our method \cite{mankin4} of higher-order Green's
functions we have developed a new approach to the electromagnetic
radiation on a curved spacetime. In a recent communication
\cite{mankin9}
we presented some initial results obtained within the framework of this
approach. Namely, we considered a pulsed source of electromagnetic
radiation in arbitrary bounded motion in a weak gravitational field
and concluded that generally the received radiation tail arrives
after a time delay which represents geometrical backscattering by
the central gravitational source. This delay effect of the wave
tails may be of great importance by their observational detection.
Further, by applying this approach to a compact astrophysical binary
system we demonstrated that under certain conditions the tail energy
can be a noticeable fraction of the direct pulse energy. 
The underlying formulas involved are herein published for the first time.

The wider aim of the present paper is to provide a comprehensive
view of our new approach to the electromagnetic radiation, more
fully describe the already published results and expand upon them.  
Specifically, among the rest we will prove the following.
(i) Vanishing of the Ricci
curvature tensor is the necessary and sufficient condition for the
validity of the Huygens' principle in the first-order approximation,
as it is already known.
(ii) If the direct pulse of electromagnetic radiation has passed the
gravitational source, then in the first approximation the structure
of electromagnetic wave tails is independent of the higher multipole
moments of the gravitational source, including the angular momentum. 
(iii) In an arbitrary weak  gravitational field it is valid that
in the first approximation the nonlocal radiative electromagnetic
tail term at infinity acquires a universal form, viz. Eq. (\ref{45}),
which is independent of the multipole order.
(The limitations of applicability as well as the
relationships of these results with the earlier ones will be
discussed in due course below.)

Finally, we think that our method deserves to be presented in detail
as it may prove to be useful also for
the theoretical investigations of detection of gravitational waves
performed within the framework of the LISA mission.
 
The remainder of the paper is organized as follows. Sec. II 
gives  a review of the theory of classical 
and higher-order Green's functions (fundamental solutions) for vector 
wave equations, as well as the recurrent formulas for calculating 
the Green's functions proceeding from the Hadamard coefficients. 
In Sec. III we determine the multipole moments of electromagnetic 
field source term with respect to a given worldline, and  
also present an algorithm for calculating the exact multipole 
solutions of the wave equations. 
In Sec. IV the main results are obtained. We consider the tail term 
of the retarded Green's function expanded to first order in the 
gravitational potential, and give the first-order tail term for the 
multipole solutions of electromagnetic field (IV A, B). In Sec. IV D 
we turn to nonlocal radiative wave-propagation correction in the 
far wave zone. We estimate the magnitude of electromagnetic radiation 
energy and find that,compared to the energy of the direct pulse,
the value of tail energy can be
considerable  and may thus have 
astrophysical significance. Limitations to the applicability of the derived 
conclusions are also considered. 
Sec. V contains brief concluding remarks. Appendix explains our notation
and displays the relevant definitions.

\section{METHOD OF HIGHER-ORDER GREEN'S FUNCTIONS}

To investigate the electromagnetic wave tails within the framework
of general relativity, we first consider on a pseudo-Riemannian
4-space $M$ a vector wave equation
\be
 ${\bf L u =f}$,\label{1}
\ee
which in local coordinates can be written in the following coordinate
invariant form (for our notation see Appendix)
\be
 L u_c := g^{ab} \nabla_a \nabla_b u_c - R^a_c u_a = f_c,
\eqnum{1'}\label{1'}
\ee
where the contravariant components of the metric tensor $g^{ab}$ are 
assumed to be of differentiability class $C^{\infty}$, and $R_{ac}$
are the covariant components of the Ricci tensor. The inhomogeneous
term $\bf f$ of Eq. (\ref{1}) is in general a distribution, i.e.
${\bf f} \in {\cal D}'^1 (\Omega)$.
In order to be able to complete the construction of ${\bf u}$, we
restrict the solutions to a causal domain $\Omega \subseteq M$ (see
point 5 in Appendix and Refs. \cite{gynther,fried}). 

The notation and basic definitions used in this paper will be in
detail presented in Appendix. Here we touch on only the spacetime
subdomains frequently resorted to. 
$C^+ (y)$ is the future light cone, i.e. the set of all points $x \in
\Omega$ that can be reached along future-directed null geodesics from
$y$; $C^- (y)$ is the past light cone, defined similarly by
past-directed geodesics. The sets $D^{\pm}(y)$ denote the respective
interiors of the future and past light cones $C^{\pm} (y)$, whereas
$J^{\pm} (y) := D^{\pm} (y) \cup C^{\pm} (y)$.

\subsection{Classical (zeroth-order) Green's function}

{\it The classical (i.e. zeroth-order) Green's function}, or
fundamental
solution as mathematicians would say, ${\bf G} (x,y)$ of the wave
equation
(\ref{1}) satisfies
\be
 {\bf L G} (x,y) = {\bf g} (x,y) \de (x,y), \label{3}
\ee
where $\de (x,y)$ is the Dirac delta distribution, with $(\de (x,y), 
\phi (x)):= \phi (y)$ for all $\phi \in C_0^{\infty} (\Omega)$ and 
${\bf g} (x,y)$ is a transport bivector (for its defining equations
see Appendix). 

As in the case of flat spacetime, there are two particularly
important Green's functions of the wave equation (\ref{1}):
the retarded Green's function ${\bf G}^+ (x,y)$ and the advanced
Green's function ${\bf G}^- (x,y)$. It has been demonstrated that
they have the following form  \cite{gynther,fried}:
\be
 {\bf G}^{\pm} (x,y) = {1 \over 2 \pi} \left[ {\bf U} (x,y)
\de_{\pm}^{(0)}
 (\si (x,y))
 + {\bf V} (x,y) \Theta_{\pm} (\si(x,y)) \right],
\label{8}
\ee
where the bitensors ${\bf U} \in C^{\infty}(\Omega\times\Omega)$ and
${\bf V}\in C^{\infty} (\Omega \times \Omega)$ are the Hadamard
coefficients of the classical Green's functions (\ref{8}) of the
vector wave equation (\ref{1}).

The bitensor ${\bf U}$ in Eq. (\ref{8}) is determined by the
following transport equation and normalization condition: 
\bea
 \si^{;a} (x,y) \nabla_a U^i_b (x,y) +M(x,y) U^i_b (x,y) =0,\nonumber\\
U^i_a (x,x)=\delta^i_a, \label{8-1}
\eea
where
\be
 M(x,y):= {1 \over 2} \nabla^a \nabla_a \si (x,y) -4.\label{mxy}
\ee
It is well known that the bivector ${\bf U}$ can be written as
\cite{fried,dewitt}
\be
{\bf U}:={\bf g}(x,y) k(x,y), \label{8-2}
\ee
where ${\bf g}(x,y)$ and $k(x,y)$ are respectively the transport
bivector and scalarized van Vleck determinant, defined in Appendix.

The bivector ${\bf V}\in C^{\infty} (\Omega \times \Omega)$ in Eq.
(\ref{8}), called the tail term, is determined by the characteristic
Cauchy problem. It corresponds to the 'logarithmic term' of the
Hadamard construction and is inherently connected with the concept
and validity of Huygens' principle \cite{hadamard}. In the regions
$D^{\pm}(y)$ the vector field ${\bf V}$ satisfies the homogeneous
wave equation
\be
 {\bf LV} (x,y) =0, \label{9}
\ee
which is completed by the characteristic initial conditions
\be
 \si^{;a} (x,y) \nabla_a V^i_b (x,y) +(M(x,y) +2) V^i_b (x,y) =
-{1 \over  2} LU^i_b (x,y), \label{10}
\ee
for $\forall x \in C^{\pm} (y)$.

Sometimes, instead of solving the differential equations (\ref{9})
and (\ref{10}) straightforwardly, in order to find the tail term
${\bf V}$, it is preferable to use an exact integral equation.
We proceed from the fact that Friedlander has derived the
corresponding exact integral equation for the tail term of the
Green's function of the scalar wave equation (Eq. (5.4.19) in Ref.
\cite{fried}), and suggested a procedure for obtaining the tensor
field Green's functions from the scalar Green's
function. Thus, generalizing the above-mentioned Friedlander's
equation, we find for the vector case when $x \in D^+ (y)$,
the tail term ${\bf V}$ satisfies the following exact integral
equation:
\be
 V_a^i (x,y) + {1 \over 2 \pi} \int_{\Sigma (y)} V^p_a (x,z)
LU^i_p (z,y)\mu_{\si (z,y)} (z) + {1 \over 2 \pi} \int_{S(y)}
U^p_a (x,z) LU^i_p (z,y) \omega (z)= 0 , \label{10-1}
\ee
Here: the $3$-surface $\Sigma(y)$ and the $2$-surface $S(y)$ are
defined by $\Sigma (y):=C^+ (y) \cap J^- (x)$ and
$S(y):= C^+ (y) \cap C^- (x)$, respectively; the operator
${\bf L}$ acts at $z$; the $3$-form $\mu_{\si (z,y)}$ and
the $2$-form $\omega$ are defined in Appendix by  Eqs. (\ref{6}) and
(\ref{6-1}). 

\begin{figure}
\centerline{
\psfig{file=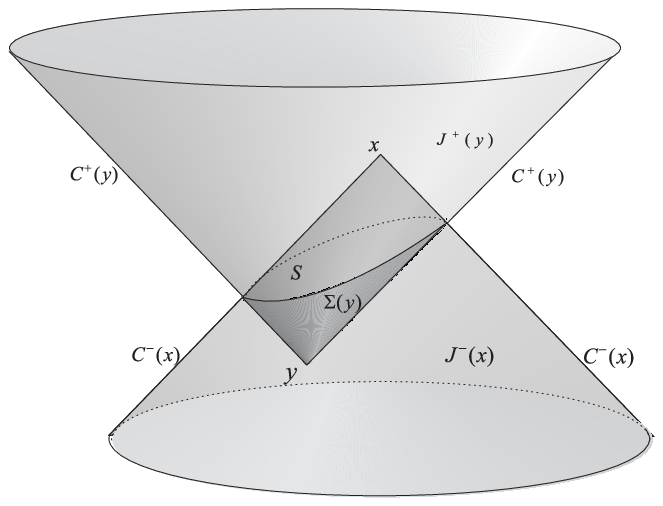,width=85mm} }
\caption{The geometric regions occurring in Eq. (9):
an illustration on a 3-dimensional intersection of the Minkowski
space with the plane $x^3={\text const}$. The darker region on
the future light cone $C^+ (y)$ is the hypersurface $\Sigma (y)$
whose $2$-dimensional boundary $S$ is represented by an ellipse
which is the intersection of the light cones $C^+ (y)$
and $C^- (x)$.
}
\end{figure}


We will use Eq. (\ref{10-1}) in Sec. IV for obtaining an expression
for electromagnetic wave tails in the first-order approximation. 

Only the retarded Green's functions ${\bf G}^+$ will be discussed, as
the corresponding results for the advanced Green's functions ${\bf G}^-$
can be obtained by reversing the time orientation on the domain $\Omega$.

\subsection{Higher-order Green's functions for vector wave equations}

Let $\Omega$ be a causal domain. Then the tensor differential
operator ${\bf L}$ in (\ref{1}) has a unique retarded $\mu$th-order
Green's function ${\bf G}^+$ on $\Omega$ such that 
\[
 LG^{+a}_{i J(\mu)} (x,y) =(-1)^{\mu} \nabla_{B(\mu)}
\left[ g^a_i (x,y) g^{B(\mu)}_{J(\mu)} (x,y)
 \de (x,y)\right], 
\]
\[
\text{supp} ~ {\bf G}^+ \subset J^+ (y).
\]
If test vector fields ${\bf \Phi} \in {\cal D}^1 (\Omega)$ are such
that 
$\text{supp}~{\bf \Phi} \subset \Omega \backslash \{y\}$, then the
retarded $\mu$th-order Green's function ${\bf G}^+$ is of the
following form:
\be
 {\bf G}^+ ={1 \over 2 \pi} \left[ \sum\limits_{\nu=0}^{\mu} 
 {\bf W}_{\nu}^{\mu} \de_+^{(\mu-\nu)} (\si) + {\bf W}_{\mu+1}^{\mu}
\Theta_+
 (\si) \right], \label{11}
\ee
where ${\bf W}_{\nu}^{\mu}, \nu =0,\dots \mu+1$ are bitensor fields
of rank 
$1$ at $x$ and of rank $\mu+1$ at $y$ recursively determined by
\bea
 W^{\mu~a}_{\nu~i~J(\mu)} =\si_{;j_{\mu}} W^{\mu-1~a}_{\nu~i~J(\mu-
1)} +
 W^{\mu-1~a}_{\nu-1~i~J(\mu-1);j_{\mu}},\nonumber \\ 
 {\bf W}^{\mu}_{-1} = {\bf W}^{\mu}_{\mu+2}:=0, \;\quad {\bf W}^0_0 := 
 {\bf U}, \;\quad {\bf W}^0_1:={\bf V}, \label{12}
\eea
where $\nu=0,\dots,\mu+1$.

This statement was first proposed by us without proof in \cite{laas},
its proof can be found in \cite{mankin3}. Eqs. (\ref{12}) provide a
simple recurrent algorithm for calculating the $\mu$th-order Green's
functions for the vector wave equation (\ref{1}) proceeding from the
Hadamard coefficients ${\bf U}$ and ${\bf V}$ of classical Green's
function (\ref{8}). The formulas (\ref{11}-\ref{12}) constitute
the main mathematical tool which allows construction of our exact
multipole solutions of vector wave equations.

\section{EXACT SOLUTIONS}

\subsection{Multipole expansion of the electromagnetic field source}

On the basis of Dixon's ideas \cite{dixon} we will now construct a
multipole expansion of the electromagnetic field source ${\bf f}$
as follows. We first choose a unique timelike worldline $\xi$ lying
inside the worldtube $\Gamma \subset \Omega$ of the source of the
electromagnetic field that represents its dynamical properties. Such
a curve can be given as a $C^{\infty}$ embedding
$t \to y(t) \in \xi$ of an open interval $I \subset {\cal R}$ into
$\Omega$, where ${\cal R}$ is the real line. We set $v^i (t) =
{dy^i \over dt}$; this vector is assumed to be timelike and
future-directed, and it is convenient to normalize the 
parametrization so that $t$ is the proper time, which means that
$v^i v_i 
=1$. For $y(t) \in \xi$ we define $\Sigma (t) := \{x|v^i (t) \si_{;i} 
(x,y(t))=0\}$, the spacelike hypersurface consisting of all geodesics 
through $y(t)$ orthogonal to ${\bf v}(t)$. We suppose that there
exists a $3$-form $\mu_{\Sigma}$ such that $\mu (x) = d \eta \wedge
\mu_{\Sigma}$ on $\text{supp}~{\bf f}$; here $\eta (x)$ is 
defined by $v^i (\eta (x)) \si_{;i} (x,y(\eta(x)))=0$, 
$y(\eta) \in \xi$. For the sake of simplicity, 
we assume also that $\text{supp}~{\bf f}$ is compact in the domain
$\Omega$. Regarding the source function ${\bf f}$ as a regular 
distribution with compact support, we can write
\[
 ({\bf f},{\bf \Phi})=\int_I dt \int_{\Sigma (t)} <{\bf f}(x) 
 {\bf \Phi} (x)> \mu_{\Sigma} (x),
\]
where $<{\bf f \Phi}>$ denotes the scalar product.
Let ${\bf \bar M}_\mu (t)$ be $C^{\infty}$ tensor fields of
corresponding 
ranks $\mu+1 \geq 1$ at $y(t) \in \xi$, with $\text{supp}~{\bf \bar
M}_\mu (t)$ 
compact, and let $N \geq 0$ be a integer. We consider the line 
distributions
\be
 f^a_N = \sum\limits_{\mu=0}^N (-1)^\mu \bar M_\mu^{I(\mu)j} (t)
\nabla_{A(\mu)} 
 g^{A(\mu)}_{I(\mu)} (x,y(t))g^a_j (x,y(t)) \bar \de (x,\xi) \in {\cal
E}'^1 
 (\Omega)   \label{13}
\ee
which assigns to any ${\bf \Phi} \in C_0^{\infty} (\Omega)$ the
number 
\be
 ({\bf f}_N, {\bf \Phi}) := \sum\limits_{\mu=0}^N \int_I \bar
M_\mu^{I(\mu)j} (t) 
 \phi_{j;I(\mu)} (y(t)) dt. \label{14}
\ee
In the particular case of $N=0$ the formula (\ref{14}) means that 
\[
 (g^a_i (x,y(t)) \bar M^i_0 (t) \bar \de (x,\xi), \phi_a (x)):=\int_I 
 \bar M_0^i (t) \phi_i (y(t)) dt.
\]

Let us now choose a test function so that $\text{supp}~{\bf \Phi}
\supset 
\text{supp}~{\bf f}$. If for all test functions such that
$\text{supp}~
{\bf \Phi} \supset \text{supp}~{\bf f}$, the following equation is
valid
\be
 ({\bf f}_N, {\bf \Phi})=\int_I dt \int_{\Sigma (t)} <{\bf f} (x)
{\bf 
 \Phi}_N (x,t)> \mu_{\Sigma} (x), \label{15}
\ee
where ${\bf \Phi}_N$ is determined by the Taylor expansion of the 
vectorfunction ${\bf \Phi}$
\[
 \phi^a_N (x,t):= \sum\limits_{\mu=0}^N {(-\mu)^l \over 2^\mu \mu!} g^{aj}
(x,y(t))
 \si^{;i_1} (x,y(t)) \dots \si^{;i_\mu} (x,y(t)) \phi_{j;I(\mu)} (y(t)),
\]
then the line distribution ${\bf f}_N$ is called the 
{\it $N$th-order multipole expansion of the source function ${\bf
f}$}.

If ${\bf f}_N$ is the $N$th-order multipole expansion of ${\bf f}$,
then it 
follows form (\ref{14}) and (\ref{15}) that one can choose
\be
 \bar M_\mu^{I(\mu)j} (t) = {(-1)^\mu \over 2^\mu \mu!} \int_{\Sigma (t)} g^j_a 
 (x,y(t)) f^a (x) \si^{;i_1} (x,y(t)) \dots \si^{;i_\mu} (x,y(t)) 
 \mu_{\Sigma} (x). \label{16}
\ee

The tensor field ${\bf \bar M}_\mu (t)$ determined by the expression
(\ref{16}) 
is called the 
{\it $2^\mu$-pole moment of the field source ${\bf f}$ with respect to
$\xi$}.
One should mention that Eqs. (\ref{14}) and (\ref{15}) do not uniquely
determine the structure of the multipole moments ${\bf \bar M}_\mu$ in
Eq. (\ref{16}). For example, let us define a set of new multipole moments
${\bf M}_\mu (t)$ by relations $M_\mu^{I(\mu)j} (t) = \bar M_\mu^{I(\mu)j} 
(t) + D_{\mu-1}^{I(\mu-1)j} v^{i_\mu} + {\de \over \de t} 
D_\mu^{I(\mu)j} (t) \;$,
$\mu \leq N \;$, $D_N^{I(N)j}=0\;$, where the symbol ${\de \over \de t}$
denotes the absolute derivative along the worldline $\xi$. Now,
integrating Eq. (\ref{14}) by parts, it follows that ${\bf f}_N$ remains 
unchanged for arbitrary tensor fields ${\bf D}_\mu (t)$ with compact 
support. 

The multipole moments defined by Eq. (\ref{16}) are obviously
symmetric in the first $\mu$ indices, i.e. $\bar M_\mu^{i_1 \dots i_\mu j}
= \bar M_\mu^{(i_1 \dots i_\mu)j}$.  They also satisfy the orthogonality
condition $\bar M_\mu^{i_1 \dots i_\mu j} v_{i_\mu}=0$. One should say,
though, that due to charge conservation, $\nabla_a f^a =0$, the
multipole moments of different orders ${\bf \bar M}_\mu$ are
interrelated through differential constraints. For that reason their
direct usage in solving particular physical problems is inconvenient.
However, Dixon demonstrated in Ref. \cite{dixon} that by means of the
multipole moments ${\bf \bar M}_\mu$ a new set of reduced multipole
moments ${\bf Q}_\mu\;$, $\mu \geq 1$ can be constructed, which together
with the total charge $Q$ will completely describe the four-current
${\bf f}$. If  $Q=\text{const}$, then the charge is automatically
conserved, and thus the multipole moments of different orders
${\bf Q}_\mu\;$, $\mu \geq 1$ are independent. Dixon's reduced multipole
moments have the following symmetry and orthogonality properties:
\bea
 Q_{\mu+1}^{i_1 \dots i_\mu jk} = Q_{\mu+1}^{(i_1 \dots i_\mu)[jk]} \quad 
 \text{for} \quad \mu \geq 0, \nonumber \\
 Q_{\mu+1}^{i_1 \dots i_{\mu-1}[i_\mu jk]}=0, \quad Q_{\mu+1}^{i_1 \dots i_\mu
jk} v_{i_\mu} =0 \quad \text{for} \quad \mu \geq 1. \label{17}
\eea

Following Dixon's work \cite{dixon}, we will in this paper
approximate the source of electromagnetic waves ${\bf f}$ by its
$N$th-order multipole expansion (\ref{13}) replacing, however,
the multipole moments ${\bf \bar M}_\mu (t)$ with the reduced ones
${\bf M}_\mu (t)$:
\be
 M_\mu^{I(\mu)j}:={2\mu \over \mu+1} Q^{i_1 i_2 \dots i_{\mu-1} i_1 j}, 
 \qquad \mu \geq 1. \label{18}
\ee

\subsection{Exact solutions of vector wave equations with a multipole 
source term}

In one of our earlier papers \cite{mankin4} we demonstrated how 
to obtain a solution by way of describing $2^{\mu}$-pole radiation
of a vector field in terms of the higher-order Green's functions, 
i.e. how to calculate the retarded solution ${\bf u}^+_{\mu}$ of the 
vector wave equation
\[
L u^a_{\mu} = \rho^a_{\mu}
\]
with a multipole source term
\be
\rho^a_{\mu}:=(-1)^{\mu} M_{\mu}^{I(\mu)j} (t)
\nabla_{A(\mu)}
 (g^{A(\mu)}_{I(\mu)} (x,y(t)) g^a_j (x,y(t)) \bar \de (x,\xi)).
\label{19}
\ee
Here ${\bf M}_{\mu} (t)$ is a tensor field of order $\mu+1$ on the
worldline $\xi$ of the source of (electromagnetic) radiation and the
coordinates of the points of the line $\xi$ are denoted by $y(t)$,
where the parameter $t$ is the proper time along $\xi$.
Let us suppose that there is a $t_0 \in I$ such that
${\bf M}_{\mu} (t) =0$ for $t < t_0$. If ${\bf M}_{\mu}$ is
determined by relations (\ref{18}), it follows from the symmetry
properties of Dixon's moments(\ref{17}) that
$\nabla_a \rho^a_{\mu} =0$, i.e.
\be
 (\nabla_a \rho^a_{\mu}, \phi) =-(\rho^a_{\mu}, \nabla_a \phi) =0.
\label{20}
\ee

On the domain $J^+ (\xi) \backslash \{\xi \}$, with $J^+ (\xi) := 
\bigcup_{y \in \xi} J^+ (y)$, the solution ${\bf u}^+_{\mu}$ can be 
represented as a regular distribution (function). To find a form of
the solution suitable for applications it is reasonable to define a
new, retarded time coordinate $\tau (x)$ as follows
\be
 \si (x,y(\tau(x)))=0, \qquad y(\tau) \in C^- (x). \label{21}
\ee
Evidently, the future light cone $C^+ (y(t))$ is determined by the 
equation $\tau (x) = t$. We denote the corresponding Leray form by 
$\mu_{\tau} (x)$, i.e. $d \tau \wedge \mu_{\tau} (x) = \mu (x)$ on
$J^+ (\xi) \backslash \{\xi\}$, (see Ref. \cite{fried}). On the
surface $C^+(y(t)) \backslash \{y(t)\}$ we have 
\be
 d \si = -\si_{;i} (x,y(t)) v^i (t)|_{\tau (x)=t} d \tau. \label{22}
\ee
In what follows we shall use the notation 
\be
 \psi (x,t):=-\si_{;i} (x,y(t)) v^i (t).  \label{23}
\ee

Due to the theorem 5 proved in \cite{mankin4}, the unique retarded
solution ${\bf u}^+_{\mu}$ of Eq. (\ref{19}) can be written in the
following form:
\bea
 u^{+a}_{\mu} (x) ={1 \over 2 \pi} \sum\limits_{\nu=0}^{\mu} \left[
\left({1 \over \psi (x,t)} {d \over dt} \right)^{\mu-\nu}
{M_{\mu}^{I(\mu)j}(t)W^{\mu~a}_{\nu~j~I(\mu)} (x,y(t)) \over
\psi (x,t)} \right]_{t= \tau(x)} \nonumber \\
 + {1 \over 2 \pi} \int_{t_0}^{\tau(x)} V^a_{j;I(\mu)} (x,y(t)) 
 M_{\mu}^{I(\mu)j} (t) dt, \qquad \forall x \in J^+ (\xi) \backslash 
 \{\xi\}, \label{24}
\eea
where the quantities ${\bf W}_{\nu}^{\mu}$, $\nu=0,\dots,\mu$ are
bitensor fields of rank $1$ at $x$ and of rank $\mu+1$ at $y$
recursively determined by Eqs. (\ref{12}), and the bitensor field
${\bf V}$ is the tail term of the Green's function of Eq. (\ref{1}).
Identifying the tensor field ${\bf M}_{\mu}$ in
Eqs. (\ref{19}) and (\ref{24}) with the tensor field (\ref{18}),
defined via Dixon's multipole moments, it is easy to see that due
to Eq. (\ref{20}), and the uniqueness of the retarded solution of
the wave equation, the solution $u^{+a}_{\mu}$ satisfies the gauge
condition $\nabla_a u^{+a}_{\mu} =0$. Consequently, the quantity
$u^{+a}_{\mu}$ can be interpreted as the retarded potential of
electromagnetic field. Thus Eq. (\ref{24}) with the recurrence
system (\ref{12}) enables one to find with admirable ease the exact
multipole solution of arbitrary order for the electromagnetic wave
equation (\ref{19}) by means of the world function $\si$, transport
bitensor ${\bf g}$ and tail term ${\bf V}$ of the classical Green's
function (\ref{8}).

\section{ELECTROMAGNETIC WAVE TAILS IN CASE OF A WEAK GRAVITATIONAL
FIELD}

Application of formulas (\ref{24}) presupposes knowledge of the tail
term ${\bf V}$ of the classical Green's function and the world
function $\si$, the exact
forms of which, however, have been thus far calculated for some
particular cases only, such as for example the Bianchi-type I metric
\cite{nariai}, de Sitter metric \cite{fried} and a class of Robertson-Walker 
metrics \cite{mankin5}. Nevertheless, for most cases approximate forms of
${\bf V}$  and $\si$ are known and can be used within the framework of our
formalism in obtaining approximate solutions. 

We assume that the gravitational field is weak and therefore use a 
perturbational approach. The metric tensor and other geometrical 
quantities are supposed to depend on a small parameter $\vaps$ which
determines the order
of deviation from flat spacetime. Such expansion has been used for
instance for the electromagnetic field in Ref. \cite{mankin6} and for
gravitational waves in Ref. \cite{thorne}.

\subsection{First-order tail term for the Green's function}

We assume that the metric tensor $g_{ab}$ is expanded up to the first
order in $\vaps$, so that the zeroth-order term is the metric tensor
of the flat spacetime $\!\!\!\!{\mathop{\;\;\; g_{ab}}\limits_0^{}}$,
i.e.
\bea
 g_{ab} =\!\!\!\!{\mathop{\;\;\; g_{ab}}\limits_0^{}} + \vaps
\gamma_{ab}
 + O(\vaps^2), \nonumber\\ 
 g^{ab} =\!\!\!\!{\mathop{\;\;\; g^{ab}}\limits_0^{}} -
\vaps \gamma^{ab} + O(\vaps^2) \label{25}
\eea
with $\gamma^{ab}=\!\!\!\!{\mathop{\;\;\;g^{ac}}\limits_0^{}}\!\!\!\!
{\mathop{\;\;\; g^{bd}} \limits_0^{}}{\gamma_{cd}}$. The world
function $\si (x,y)$, the transport tensor ${\bf g} (x,y)$ and the
scalarized van Vleck determinant $k(x,y)$ can then also be expanded
in the parameter $\vaps$. We have
\bea
 \si = {\mathop{\si}\limits_0^{}} +  \vaps {\mathop{\si}\limits_1^{}}
 + O(\vaps^2), \nonumber\\ 
 {\bf g} = {\mathop{\bf g}\limits_0^{}} + \vaps {\mathop{\bf
 g}\limits_1^{}} + O(\vaps^2), \nonumber\\ 
 k = {\mathop{k}\limits_0^{}} + \vaps {\mathop{k}\limits_1^{}}
 + O(\vaps^2),\label{25-1}
\eea
where $\mathop{\si}\limits_0^{}$,
${\mathop{\bf g}\limits_0^{}}$ and
$\;\mathop{k}\limits_0^{}=1$ are the corresponding quantities of flat
spacetime. The expansions of the Ricci tensor and the tail term
${\bf V}$ of the classical Green's function begin with terms which
are small of the first order, that is
\bea
 R_{ab} =\vaps  \!\!\!\!\!{\mathop{\;\;\; R_{ab}}\limits_1^{}}
 + O(\vaps^2), \nonumber\\
 {\bf V} = \vaps {\mathop{\bf V}\limits_1^{}} 
 + O(\vaps^2). \label{26}
\eea
Due to the tensor character of the expressions obtained, they can be
used in all the coordinate systems, where the metric preserves the
form (\ref{25}). Up to now we have where necessary meticulously
inserted the underscripts $0$ and $1$ denoting the zeroth-order
terms (i.e. the flat spacetime quantities) and the first-order 
terms, respectively. The Ricci tensor $R_{ab}$ and the tail term
${\bf V}$ being small of the first order, we will omit these
underscripts where confusion is excluded, and write the results in
what follows to the first-order approximation. 

Next we will find an explicit expression for the leading member
of the expansion of the tail term ${\bf V}$ proceeding from the
integral equation (\ref{10-1}). 
Since ${\bf V}$
is small of the first order, it follows from Eq. (\ref{10}) that
the quantity $LU^i_b (x,y)$ is also small of the first order. Thus the
first integral in Eq. (\ref{10-1}) is a second order small quantity,
whereas the first factor  $U^a_p (x,z)$ under the second integral
in Eq. (\ref{10-1}) may be considered as 
a zeroth order quantity. Hence, taking into account Eq. (\ref{8-2})
with $\;\mathop{k}\limits_0^{}=1$, the vector integral equation
(\ref{10-1}) reduces to 
\be 
 V^i_a (x,y) = - {1 \over 2 \pi} \int_{S(y)} g^p_a (x,z) L U^i_p
(z,y) 
 \omega (z) + O(\vaps^2). \label{28}
\ee
Likewise, the first factor under the last integral is of zeroth
order, and if we would use Minkowskian coordinates in the flat
spacetime we could write $g^p_a=\delta^p_a + O(\vaps)$.
Nevertheless, in order to be able to use general curvilinear
coordinates in the flat spacetime without any need for modifications
in the integral in Eq. (\ref{28}), we prefer to keep the more
general expression
$g^p_a (x,z)$. The same will be done in similar cases below.   

To calculate the factor $LU^i_a$ under the last integral we need
the linear terms in the expansions of $g^i_a(x,y)$ and $k(x,y)$
which can obtained as follows. Expanding the defining equation
$\sigma^{;a} \nabla_a g^a_i (x,y)=0$ of the transport bivector, we
have
\be
g_{ai}(x,y) = \!\!\!\!{\mathop{\;\;\; g_{ai}}\limits_0^{}}
 + {\vaps \over 2} \int_0^1 
\!\!\!{\mathop{\;\; g{^p_a}}\limits_0^{}}(y,z(\la))
\!\!{\mathop{\;\; g{^q_i}}\limits_0^{}}(y,z(\la))
\left[\gamma_{q(p;r)}(z(\la))-
\gamma_{pr;q}(z(\la))\right] \dot z^r d\la
+O(\vaps^2). \label{28-1}
\ee
Here, as well as in the next integral, integration is performed along
the geodesic $z(\la)$ connecting 
the points $x$ and $y$, where $\la$ is an affine parameter on the 
geodesic, such that $z(1)=x$, $z(0)=y$. 

Next substituting Eq. (\ref{8-2}) with Eq. (\ref{28-1}) into Eq.
(\ref{8-1}) yields  for the scalarized van Vleck determinant
(\ref{8-3})
the following first-order expression
\be
 k(x,y) = 1 + {1 \over 8} \si^{;a} (x,y) \si^{;b} (x,y) \int_0^1 
\la (1-\la) g^p_a (x,z(\la)) g^q_b(x,z(\la)) R_{pq} (z(\la)) d \la
+O(\vaps^2). 
 \label{29}
\ee

Finally, after quite simple but 
voluminous calculations we obtain from Eqs. (\ref{28}-\ref{29})
the seeked expression for the first-order tail term
\be
 V^i_a (x,y) = {1 \over 4 \pi} \left\{ g^i_a \nabla_b \left[ \si^{-
1}  
 (x,y) P^b (x,y) \right] + \si^{-1} (x,y) F^i_a (x,y)
\right\},\label{30}
\ee
where
\be 
 P^b (x,y):= \int_{\Sigma (y)} g^b_p (x,z) G^{pq} (z) d \Sigma_q (z), 
 \label{30-1}\\
\ee
\be 
F^i_a (x,y):=\int_{S(y)} g^p_a g^{iq} \left( 4 R_{[p}^r D_{q]r} + R
 D_{pq} +  2 R_{pq} \si (x,y) \right) \omega (z). \label{30-2}
\ee
Here $G^{pq} := R^{pq} - {1 \over 2} g^{pq} R$ are the contravariant
components of Einstein's tensor, $R$ is the Ricci scalar, 
$D_{pq} := \si_{;[p} (z,x) \si_{;q]} (z,y)$ and $d \Sigma_q (z)$ 
denotes the $3$-surface element on 
$\Sigma (y)$ defined by
\be
 d \Sigma_p (z) = {1 \over 6} \sqrt{g(z)} \epsilon_{pqrs} dz^q
\wedge dz^r
 \wedge dz^s, \label{30-3}
\ee
where $\epsilon_{pqrs}$ are the components of the discriminant
tensor with 
$\epsilon_{0123}=1$. 

It can be shown that $V^i_a (x,y)$ remains bounded when $x \to \bar
x \in C^+ (y)$. 
From Eqs. (\ref{30}-\ref{30-2}) we see that $V^i_a$ vanishes 
for every $x \in \Omega$ and $y \in \Omega$ if $R_{pq} (z) =0$,
$\forall
z \in \Omega$. It is the necessary and sufficient condition for
validity of 
Huygens's principle in the first-order approximation 
\cite{mankin6,gynther2}.

Among the astrophysical applications of great interest is the case
in which the source of gravitation is spatially isolated, i.e. 
$\text{supp}~R_{ab} \subset \tilde \Gamma$, where $\tilde \Gamma$ is
the
world tube of the gravitational field source.
In this case it is reasonable to divide the spacetime domain   
$J^+ (y)$ into the following three subdomains:
\[
 D(y) := \{x|x \in J^+ (y), \Sigma (y) \cap \tilde \Gamma =
\emptyset \},
\]
\[
 D^* (y) :=\! \{x|x \!\in\! D^+ (y), C^+ (y) \cap \tilde \Gamma \!\subset\! 
 \Sigma (y), S(y) \cap \tilde \Gamma = \emptyset \},
\]
\be 
\tilde D := \{x|x \in J^+ (y), S(y) \cap \tilde \Gamma \not=
\emptyset \}. 
 \label{31}
\ee

It is evident from Eqs. (\ref{30}-\ref{30-2}) that ${\bf V} (x,y) =0$ 
if $\forall x \in D(y)$. We remind that our results are valid for
an arbitrary weak gravitational field. An analogous conclusion on the
background of a weak Schwarzschild field was drawn in Refs.
\cite{dewitt2,dewitt3,colleau}.
As in the first approximation $P^b_{;b} =0$, $x \in D^* (y)$, the
expression
for the tail term ${\bf V}$ in the domain $D^* (y)$ becomes
considerably
simpler, namely
\be
 {\bf V} (x,y) = {1 \over 4 \pi} {\bf g} (x,y) P^b \nabla_b \left({1
\over 
 \si (x,y)} \right). \label{32}
\ee

Thus, in the subdomains (\ref{31}) the character of the
tail term
is qualitatively different. (i) {\it No-tail region.} 
In the subdomain $D$ the observer at
$x$ does
not see any wavetail, ${\bf V} (x,y) =0$, from the source at $y$.
(ii) {\it Simple-tail region.} 
In the subdomain $D^*$ there appears a wavetail of the
simple form (\ref{32}). (iii) {\it General-tail region.} 
In the subdomain $\tilde D$
one can examine a wavetail of the general structure (\ref{30}) (see
Fig. 2).

\begin{figure}
\centerline{
\psfig{file=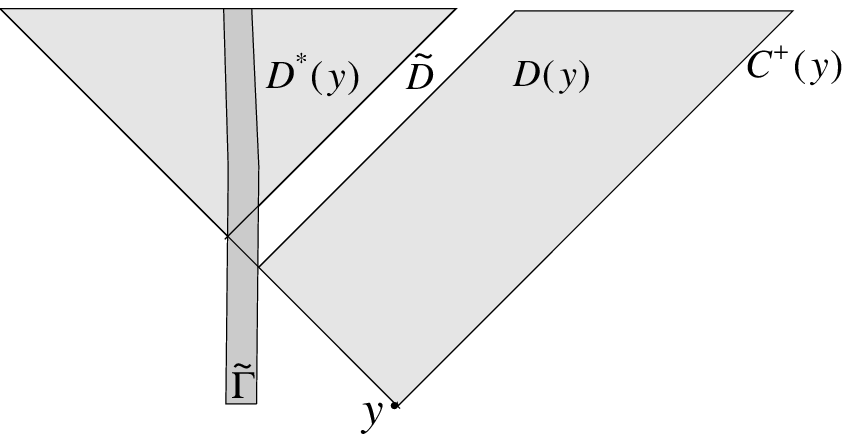,width=85mm} }
\caption{
A diagram in Minkowski 2-spacetime: the regions where the
character of the tail term is qualitatively different.
$y$ is the source of electromagnetic waves and $\tilde \Gamma$ is
the world-tube of gravitational source. The observer in the region
$D(y)$ does not see any wavetail, the observer in $D^*(y)$ can see
a wavetail of simple structure (37) whereas in the region
$\tilde D(y)$ one can see a wavetail of the general form (30). 
}
\end{figure}

Eq. (\ref{32}) leads to the conclusion that if the metric tensor
satisfies the Einstein field
equations,
the tail term of the Green's function in the first approximation
in the
spacetime domain $D^* (y)$ is completely determined by the four-
momentum of
the source of gravitation. The higher multipole moments of the source
of gravitation, including the angular momentum, do not influence 
in this approximation the structure of  
${\bf V} (x,y)$ (in the domain $D^* (y)$). A corresponding result
for a scalar
field has been presented in Ref. \cite{mankin7}, and for a vector
field in Ref.
\cite{mankin6}.

\subsection{Tail term for multipole solution}

On the basis of Eq. (\ref{24}), the tail term ${\cal V}_{\mu}^a$ of
the
retarded $2^{\mu}$-pole solution of Eq. (\ref{19}) can be written as
\be
 {\cal V}_{\mu}^a = {1 \over 2 \pi} \int_{t_0}^{\tau (x)}
V^a_{j;I(\mu)} 
 (x,y(t))M_{\mu}^{I(\mu)j} (t) dt. \label{33} 
\ee

Next we will study the tail term ${\cal V}_{\mu}^a$ of a multipole
solution in a weak gravitational field more closely. 
In a weak gravitational field the bivector field ${\bf V} (x,y)$ is
determined by Eq. (\ref{30}). 
In what follows
we presume that the worldline $\xi$ of the source of
electromagnetic waves remains outside the source of gravitational
field, whereas $y(t_0)$ and $y(t_1)$ are points on the worldline $\xi$
of the wave source, $t_0$ and $t_1$ being, respectively, the proper
time values when the source begins and terminates the emission. If the
worldline $\xi$ of the wave source lies outside the world tube
$\tilde \Gamma$ of the source of gravitation, i.e.
$\text{supp}~\xi \cap \text{supp}~\tilde \Gamma = \emptyset$, and
if $\text{supp}\, {\bf M}_{\mu} (t) \subset (t_0,t_1)$, then in
the structure of the tail term  ${\cal V}_{\mu}^a$ there appear
features similar to the case of the tail term of the fundamental
solution $\bf V$ discussed at the end of the last subsection.

To begin with, we divide the spacetime domain $J^+ (y(t_0))
\backslash \{\xi\}$ with $y(t) \in \xi$ into three subdomains:
\[
 E:=\{x|x \in J^+ (y(t_0)) \backslash \{\xi\}, \Sigma (y(t_0)) \cap
\tilde 
 \Gamma = \emptyset\}, 
\]
\bea
 E^* :=\{x|x \in J^+ (y(t_0)) \backslash \{\xi\}, C^+ (y(t_1)) \cap
\tilde\Gamma \subset \Sigma (y(t_1)),\nonumber\\
 S(y(t_1)) \cap \tilde \Gamma =
\emptyset \},\nonumber
\eea
\be 
\tilde E:= \{x|x \in J^+ (y(t_0)) \backslash \{\xi\}, x \notin E, 
 x \notin E^* \}.  \label{34}
\ee

\begin{figure}
\centerline{
\psfig{file=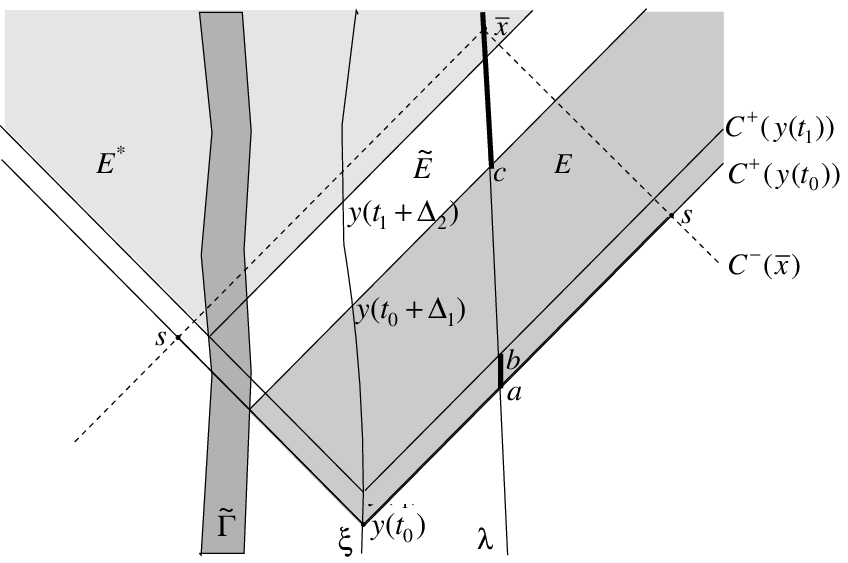,width=85mm} }
\caption{
A diagram in Minkowski 2-spacetime: the regions $E$, $E^*$
and $\tilde E$ where the tail term of a wave pulse with finite
duration behaves differently. The past light-cone $C^- (\bar x)$
originating from the point $\bar x$ is represented by the dotted
lines. The bold-faced part of the future light-cone $C^+ (y(t_0))$
corresponds to the hypersurface $\Sigma(y(t_0))$ with $x=\bar x$,
and the boundary of the hyperplane $\Sigma(y(t_0))$, i.e.,
$S(y(t_0))$ is seen as the two points $s$. $\tilde \Gamma$ is the
worldtube of the source of gravitational field. $\xi$ is the
worldline of the wave source which radiates during a finite proper
time interval $[t_0,t_1]$. $\lambda$ is the worldline of the
observer. On the observer's worldline $\lambda$ these intervals are
indicated in bold where the observer can in principle see the direct
pulse (the interval $[a,b]$) and the wave tail (the semi-interval
$[c, \infty)$). During the interval $(b,c)$ on $\tilde \xi$ there
occurs a blackout between the direct pulse and the wave tail.
}
\end{figure}

In the subdomains $E$, $E^*$ and $\tilde E$ the
character of the tail term of a wave pulse with finite duration is
qualitatively different.
(i) {\it No-tail region.}
If $x \in E$, then also $x \in D(y(t))$ for every $t \in (t_0,
\tau (x))$. 
Hence it is valid that $V^a_j =0$, and consequently
\be
 {\cal V}^a_{\mu} =0, \qquad \forall x \in E.
\ee

As $E \not= \emptyset$, {\it there exist points in space, where 
the (fore)front of the wave is not simultaneously accompanied by the
wave
tail, the latter appearing after some time delay}. To describe this
effect
in more detail, let us define the quantity $\tau_0 (x,z)$ as a
solution of
the following set of equations:
\bea
 \si (x,z) =0, \quad z \in C^- (x) \cap \tilde \Gamma, \nonumber\\ 
 \si (z, y(\tau_0 (x,z)))=0, \quad y(\tau_0) \in C^- (z). \label{36}
\eea
For further treatment are important the maximal and minimal values
of the
time interval $\tau (x) - \tau_0 (x,z)$ (measured in the proper time
of
the wave source) for the fixed spacetime point $x$, viz.
\bea
 \Delta_1 (x) := \tau (x) - \text{max}~\tau_0 (x,z), \nonumber\\
 \Delta_2 (x) := \tau (x) - \text{min}~\tau_0 (x,z). \label{37}
\eea

The quantities $\Delta_1 (x)$ and $\Delta_2 (x)$ can be given a
simple lucid interpretation by taking into account that an
instantaneous wave pulse emitted by the wave source at
$\text{max}~\tau_0 (x,z)$ travelling at the speed of light reaches
the spacetime point $x$ exactly as if it would first travel to the
source of gravitation, reflect from there and further travel to the
spacetime point $x$. In the case of $\text{min}~ \tau_0 (x,z)$ the
reflection takes place precisely when the above-mentioned
instantaneous wave pulse has passed the source of gravitation.

Obviously, in case $\tau(x) \leq t_0 + \Delta_1 (x)$, then $x \in E$
and 
${\cal V}^a_{\mu}=0$. 

(ii) {\it Simple-tail region.} 
If $\tau (x) \geq t_1 + \Delta_2 (x)$, then 
$x \in E^*$, and the structure of the tail term is determined only
by the
four-momentum of the source of gravitational field:  
\be
 {\cal V}^a_{\mu} =\! {1 \over 8 \pi^2}\!\! \int_{t_0}^{t_1}\!\!\!\! g^a_j (x,
y(t)) 
 M_{\mu}^{I(\mu)j} (t) P^b \nabla_b \!\left[{1 \over \si (x,y(t))}
 \right]_{;I(\mu)}\!\!\! dt. \label{38}
\ee

Consequently, the higher multipole moments of the source of
gravitation, 
among them the angular momentum, not influencing the structure of 
the wave tail.

(iii) {\it General-tail region.} 
To the domain $\tilde E$ corresponds the interval
$(t_0 + \Delta_1 (x), t_1 + \Delta_2 (x))$ of the retarded time
$\tau (x)$,
and for these instants it is valid
\be
 {\cal V}^a_{\mu} = {1 \over 2 \pi}
\int\limits_{t_0}^{\text{min}~(\tau (x) -
 \Delta_1 (x), t_1)} V^a_{j;I(\mu)} (x,y(t)) M_{\mu}^{I(\mu)j} (t)
 dt, \label{39}
\ee
where ${\bf V}$ is determined by Eqs. (\ref{30}-\ref{30-2}). Thus, in
comparison with the direct pulse, the tail of the wave appears to the
observer after the time delay $\Delta_1 (x)$. 

The particular case $\Delta_1 (x) > t_1 - t_0$ is also of interest.
In it
there occurs a time lapse of duration $\Delta (x)$ between the end
of the
principal pulse ($\tau (x) =t_1$) and the appearance of the wave
tail, with
\be
 \Delta (x):= \Delta_1 (x) - (t_1 - t_0). \label{40}
\ee
Hence, instead of a single pulse, the observer would see two clearly
separable pulses: the principal one and the wave tail.

\subsection{Schematic representation of the geometry of wave
tail generation}

The purpose of this subsection is to provide a schematic plane
spacetime representation of the geometry of wave tail generation.
This subsection is meant as an illustration, the rest of this
paper does not rely on it. 
For the sake of simplicity we make here the following simplyfing
assumptions which are
not used in our calculations: (i) the gravitational source is
spherical and static, and (ii) the gravitational and wave sources,
and the observer do not move with respect to each other.

\begin{figure}
\centerline{
\psfig{file=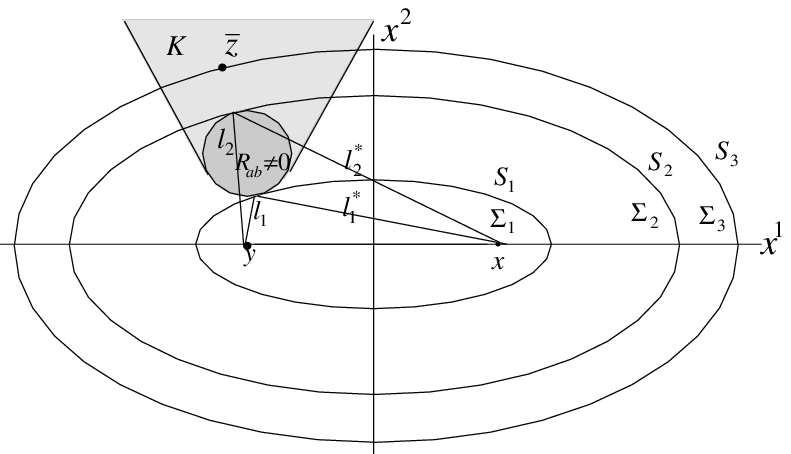,width=85mm} }
\caption{
A schematic representation of the geometry of wave tail
generation on a spacelike plane $x^1x^2$. Here $y$ is the wave
source, $x$ is the observer, the darker shadowed circular disk
is the gravitational source, and $K$ is the focusing region.
The remaining components in the figure are explained in Sec.~IV C.
}
\end{figure}

In order to draw  Fig.4 we have suppressed the time dimension and
one space dimension. Thus all the images in Fig.4 are obtained by
projecting the Minkowski spacetime onto a hypesurface
$x^0=\rm{const}$, and then cutting the 3-space with $x^1x^2$.
The plane of the figure is determined
by locations of the wave source $y$, of the observer $x$ and of the
center (not pointed out on the figure) of the gravitational source.
The surface $S(y)$, which spreads with time, is the boundary of the
ellipsoid of revolution $\Sigma(y)$, with foci at the locations of
the observer $x$ and of the wave source $y$. The ellipses $S_1$ and
$S_2$  are the intersections of the plane of the figure with the
surface $S(y)$ that, respectively, correspond to the instants of
time at which a delta-like wave pulse emitted by the wave source
reaches and passes the source of gravitation. If the surface $S(y)$
has not yet reached the gravitational source, then ${\bf V}=0$ and
there is no wave tail. If the surface $S(y)$ has passed the
gravitational source, the source will forever remain inside the
region of integration
$\Sigma (y)$, while ${\bf P}(x,y)/8\pi = \rm{const}$ will be the
\mbox{$4$-momentum} of the gravitational source and $F^i_a (x,y)=0$.
The shadowed area $K$, which includes the source of gravitational
field, corresponds to the region where the tail wave field is
predominantly generated by gravitational focusing which deforms the
direct wave fronts \cite{thorne}. Thus we have the following picture.
The wave source $y$ emits an instantaneous wave pulse. The direct
pulse propagates along the direct route $yx$ to the observer $x$,
after which there occurs a blackout before the arrival of the first
tail contribution. The parts of the wave front which travel (along
the routes $l_1$, $l_2$, etc.) to the points of the region $K$,
scatter off (reflect) from there and then propagate (along $l_1^*$,
$l_2^*$, etc.) to $x$ appearing to the observer as the tail wave.

To elucidate the introduced maximal and minimal time intervals,
$\Delta_1$ and $\Delta_2$, let us again
consider to the example presented in Fig.4. At the time
$y^0 =t_0$ the wave source emits a delta-like pulse.
The evolution of the surface $S(y)$ is characterized
by the time-varying semi-major $a$ and semi-minor $b$ axes
which depend on the observer time $x^0$ as $a=(x^0-t_0)/2$ and
$b=(\sigma (x,y))^{1/2} /2$. The ellipses $S_1, S_2$ and $S_3$ are
the intersections of the plane of the figure with the surface $S(y)$
at times of observation
$x^0_1 =t_0 + |\vec x - \vec y| + \Delta_1$, $x^0_2 = t_0 +
|\vec x - \vec y| + \Delta_2$ and $x^0_3 > x^0_2$,
respectively. Here $|\vec x - \vec y|$ is the spatial separation
between the wave source and the observer. The observer detects the
direct pulse at the time $x^0_0 =t_0 + |\vec x -\vec y|$, the wave
tail begins to appear at $x^0_1 =t_0 + l_1 + l_1^*=x^0_0 + \Delta_1$,
and beginning from the time  $x^0_2 =t_0 + l_2 + l_2^* =x^0_0 +
\Delta_2$ the structure of the wave tail is determined by Eq. (\ref{30})
with  $F^i_a =0$. The time interval
$\Delta_1 = l_1 + l_1^* - |\vec x - \vec y|$
is evidently equal to the difference of the propagation times of
two pulses from the wave source to the observer: one arriving along
the direct route and the other first travelling to the source of
gravitation, reflecting from there and then reaching the observer.

\subsection{Radiative part of the electromagnetic wave tails}

Let us examine the case in which the source of gravitation is
spatially
isolated and the distance of the source of waves from the source of
gravitation is bounded. The tail term being a first-order small
quantity,
we can regard the spacetime as flat when discussing it and use
Minkowskian coordinates whose origin lies inside the world tube of the
source of gravitation. We denote the distance of the observer to the
origin
of the coordinates by $r$, and will also below use the following
notation:
if two functions $r A(x)$ and $r B(x)$ are equal in the limit
$r \to \infty$, $\tau = \text{const}$, i.e. 
$\displaystyle \lim_{{r \to \infty \atop \tau = \text{const}}} 
(r A(x)) = \lim_{r \to \infty \atop \tau = \text{const}} (r B(x))$, 
then we will write $A(x) \doteq B(x)$. 
The position vectors of the points $x$ and $y$ in
three-space are denoted by $\vec x$ and $\vec y$, respectively. 

Taking now into account that for an arbitrary finite $\tau$ are valid
the asymptotic relations
$\displaystyle \lim_{r \to \infty} \si (x,y(t)) \sim r$,
$\displaystyle \lim_{r \to \infty} \si_{;i} (x,y(t)) \sim r$ and
$\displaystyle \lim_{r \to \infty} \si_{;ik} (x,y(t)) \sim O(1)$,
we obtain from Eqs. (\ref{30}-\ref{30-2}) the relationship
\be
 V^a_{j;I(\mu)} (x,y(t)) \doteq (-1)^{\mu} \si_{;i_1} (x,y(t)) \dots 
 \si_{;i_{\mu}} (x,y(t)) \left({1 \over \psi (x,t)} {d \over dt}
 \right)^{\mu} V^a_j (x,y(t)). \label{41}
\ee
For brevity, in what follows we will not write explicitly the
arguments 
of the functions depending both on the point $x$ as well as on the 
parameter $t$.

The radiative part of the wave tail which dominates at infinity can
be 
written as 
\be
 {\cal V}^a_{\mu~\text{rad}} \doteq {(-1)^{\mu} \over 2 \pi}
 \int\limits_{t_0}^{
 \tau(x)} \si_{;i_1} \dots \si_{;i_{\mu}} M_{\mu}^{I(\mu)j} (t)
 \left({1 \over \psi} {d \over dt} \right)^{\mu} V^a_j dt. \label{42}
\ee
Due to Eqs. (\ref{24}) ja (\ref{12}) the zeroth-order radiative part
$Z^a_{\mu}$ of the $2^{\mu}$-pole solution can be written as
\bea
 Z^a_{\mu} (x,y(\tau)) \doteq {1 \over 2 \pi} \left[ \left( {1 \over
\psi}
 {d \over dt} \right)^{\mu} g^a_j {\si_{;i_1} \dots \si_{;i_{\mu}}
\over 
 \psi} M_{\mu}^{I(\mu)j} (t) \right]_{t=\tau (x)} \nonumber \\
 \doteq {1 \over 2 \pi} g^a_j (x,y(\tau)) \left({1 \over \psi
(x,\tau)} 
 {d \over d \tau} \right)^{\mu} {\si_{;i_1} (x,y(\tau)) \dots 
 \si_{;i_{\mu}} (x,y(\tau)) \over \psi (x,\tau)} M_{\mu}^{I(\mu)j}
(\tau). 
 \label{43}
\eea
On the basis of the last two relations, after integrating by parts
the right
hand side of Eq. (\ref{42}) we obtain the following expression for
${\cal V}^a_{\mu~\text{rad}}$: 
\bea
{\cal V}^a_{\mu~\text{rad}}\doteq \int_{t_0}^{\tau(x)}\psi g^i_b V^a_i
Z^b_{\mu}(x,y(t))dt \nonumber \\
 + \left\{\sum_{\nu=0}^{\mu-1} (-1)^{\mu-\nu} \left[
 \left({1 \over \psi}{d \over dt}\right)^{\nu} V^a_j \right] 
 \left({1 \over \psi}{d \over dt} 
 \right)^{\mu-\nu-1} {\si_{;i_1} \dots \si_{;i_{\mu}} \over \psi} 
 M_{\mu}^{I(\mu)j} (t)\right\}_{t=\tau(x)}, \label{44}
\eea
where $Z^b_{\mu} (x,y(t)) := Z^b_{\mu} (x,y(\tau(x)))|_{\tau(x)=t}$. 

We must point out that Eq. (\ref{44}) contains $\mu+1$ terms, out of 
which the last $\mu$ ones are actually instantaneous. The first term 
in Eq. (\ref{44}) is truly nonlocal, whereas the factor 
$\psi g_b^i V^a_i$ assigns most of the weight to the source's recent
past.

The truly nonlocal radiative wave-propagation correction
\be
 \tilde {\cal V}^a_{\mu~\text{rad}} := \int_{t_0}^{\tau(x)} \psi
g^i_b V^a_i 
 Z^b_{\mu} (x,y(t)) dt \label{45}
\ee
takes an universal form which is independent of the multipole order.
An analogous 
result in a weak Schwarzschild field in a slow-motion approximation 
has been earlier found in Ref. \cite{leonard} (cf. also
\cite{mankin8}).
Our formula (\ref{45}) generalizes this result in two ways.
First, Eq. (\ref{45}) is valid in case of an arbitrary weak
gravitational
field in the corresponding asymptotically flat spacetime. Second,
there are
no formal restrictions to the bounded motion of the wave source, e.g.
the wave source can move with a relativistic speed.
Due to the universal form of
the tail term, Eq. (\ref{45}) has a simple, but from the point view of
observations essential, generalization.
On the assumption that the local part $A^a (x,\tau)$ of a wave
observable at infinity
(or its zeroth-order radiative part) can be
approximated
with sufficient accuracy by the superposition of a finite number of
multipole
waves, it follows from Eq. (\ref{45}) that the corresponding nonlocal
radiative wave-propagation correction $E^a (x,\tau)$ has the
following form:
\be
 E^a (x,\tau) = \int_{t_0}^{\tau (x)} \psi g^i_b A^b (x,t) V^a_i dt, 
 \label{46}
\ee
where $\psi$ is defined by Eq. (\ref{23}).

For a source radiating in the pulsed mode the last formula enables
us to comparatively simply (on the basis of the parameters of a
recorded direct pulse and a presumable model of an astrophysical
object) predict the physical parameters of the radiation tail and
evaluate the possibilities of observational detection of the tail.

It should be emphasized that applicability of Eq. (\ref{46})
does not depend on whether the source of radiation actually emits
in the pulsed mode or pulses of radiation are present at the location
of an observer due to the kinematics of the radiating system (e.g.
the rotation of the directed radiation cone of a pulsar).

If $A^b (x,\tau)$ does not vanish merely during a finite time
interval 
$t_0 < \tau < t_1$ and the worldline of the source of radiation lies
outside
the world-tube of the source of gravitation, then the conclusions of
Sec. IV B
are also valid for the radiative part $E^a$ of the tail term. Thus,
for example, if $\tau (x) \leq t_0 + \Delta_1 (x)$, then
$E^a(x,\tau) =0$; if  
$\tau (x) \geq t_1 + \Delta_2 (x)$, then
\[
 E^a(x,\tau)\!\doteq\!{1 \over 4 \pi}\!\int_{t_0}^{t_1}\!\!\!\!A^a (x,t)
\psi (x,y(t)) P^b \nabla_b\! \left({ 1 \over \si
(x,y(t)\!)}\right)dt
\]
\be
 \;\;\doteq
 {1 \over 4 \pi} \int_{t_0}^{t_1} {\si_{;b} (x,y(t)) P^b \over \si 
 (x,y(t))} {d \over dt} A^a(x,t) dt.  \label{47}
\ee

A matter of particular interest is the situation in which
$\Delta_1 (x) > (t_1 -t_0)$, as in this case there is a blackout
during
the time interval $\Delta (x) = \Delta_1 (x) - (t_1 -t_0)$ between
the end of
the direct pulse and the appearance of the wave tail. The last fact
considerably simplifies the observation methods for distinguishing
the profile
of the direct pulse from the general relativistic radiation tail
originating
from compact astrophysical binary systems. The relative intensity of
the direct
pulse and the wave tail, and the time delay (blackout) between them
can yield
essential additional information, independently of other methods,
about the
physical characteristics of a binary (the distance between the
compact objects,
the mass of the source of gravitation, the orientation of the plane
of the
orbit with respect to the observer, etc.). 

From Eq. (\ref{47}) it follows that the wave tail effect of the
astrophysical
systems radiating in the pulsed mode is predominatly caused by the
low-frequency modes ($\omega \alt 2 \pi/ (t_1 -t_0)$) of the direct 
pulse. (Here the angular frequency $\omega$ 
of the radiation is measured in the proper time of the wave source.) 
The high-frequency waves reflecting from
the spacetime curvature interfere and prevalently attenuate each
other in
the expression of the tail. Also important, from the point of view
of observational
detection, is the fact ensuing from Eq. (\ref{46})
that the intensity and the time delay of the wave tail of radiation
from a pulsed source moving on a circular orbit depend substantially on the
mutual positions of the observer, the wave source and the source of
gravitation. By comparing the profiles of the pulses emitted in different 
points of an orbit, one can distinguish in principle the contributions 
of the direct pulse and the
tail even if there is no considerable time delay between the tail
and the principal pulse. This circumstance is significant because in the
case of most astrophysically realistic models the physical conditions
necessary for the occurrence of a blackout in between the direct pulse and 
the wave tail will considerably decrease the intensity of the tail.

Let us now turn to the magnitude of electromagnetic radiative energy
arriving after the direct pulse has passed, $\tau (x) > t_1 + \Delta_2 (x)$.
The power ${d P \over d \Omega}$ radiated into the solid angle $d
\Omega$ can be calculated as follows 
\be
 I (\tau) \equiv {d P \over d \Omega} = -{r^2 \over 4 \pi} E^a_{,0} 
 (x,\tau) E_{a,0} (x,\tau). \label{48}
\ee
For further treatment it is reasonable to present Eq. (\ref{47}) in
a somewhat different form. In the proper reference frame of the
source of gravitational field we have
\be
 E^a_{,0} \doteq -2 M \int_{u(t_0)}^{u(t_1)} {[A^a_{,0} (x,\tau(x))]_{
 \tau (x)=t(u)} \over (u(\tau)-u)^2} du, \label{49}
\ee
where $u(t):=y^0 (t) - \vec n \cdot \vec y (t)$, $\vec n := \vec x / 
|\vec x|$ and $M$ is the mass of the source of gravitational field.
Below we will use notations $T:= u(t_1) - u(t_0)$, which is the duration
of the direct pulse in the observer time, and
$\tilde \Delta_2 := u(t_1 + \Delta_2 (x)) - u(t_1)$.

To get an idea of the magnitudes involved we construct an artificial
example in which
\be
 A^a_{,0}(x,\tau(x))|_{\tau=t(u)} = {1 \over r} f^a (\upsilon,\phi) 
 \Theta (u\!-u(t_0)) \Theta (u(t_1)\!-u). \label{50}
\ee
Here $\Theta$ is the Heaviside distribution, $\upsilon, \phi$ are the
polar angles determining the observer's position and $f^a$ is an
arbitrary
vector function. From Eqs. (\ref{48}-\ref{50}) we obtain an estimate
of the
ratio
of the intensity of the wave tail $I (t_1 + \Delta_2 (x))$ to the
intensity of the direct pulse $I_0$ (the ratio of the densities of
the 
energy fluxes at the observer's location), namely
\be
 {I (t_1 + \Delta_2 (x)) \over I_0} = \left({2 M \over \tilde 
 \Delta_2 (x)} \right)^2 \left({T \over T+\tilde \Delta_2 (x)}
\right)^2.
 \label{51}
\ee
We see that if $\tilde \Delta_2 (x)$ is sufficiently small $(\tilde 
\Delta_2 (x) \sim 2M)$, the intensity
of the tail can be comparable with that of the direct pulse.  
Evidently, in this case we cannot confine
ourselves to the first approximation but must instead additionally
take into consideration the higher approximation(s).

To illustrate the above estimate, let us consider a wave source
rotating in a circular orbit (of radius $r_0$) around a spherical source
(of radius $r_s$) of gravitational field. The origin of the spatial
coordinates is taken to coincide with the center of the source of 
gravitation. Under the conditions $\vec n \cdot \vec y (t_1)<0$ and $r_0 
\gg d+r_s$,
where $d:= r_0 \sqrt{1-(\vec n \cdot \vec y (t_1) / r_0)^2}$, it
follows from Eqs. (\ref{36}) and (\ref{37}) that
\be
 \tilde \Delta_2 (x) \approx {(d+r_s)^2 \over 2 r_0} + O(M).
\label{52}
\ee
Hence, if $r_0 > r^2_s /4M$, then in case the wave pulse is emitted in
the region of the geometric shadow of the source of gravitation or
in its vicinity it is valid that $\tilde \Delta_2 (x) \sim 2M$, and
according to the estimate (\ref{51}) the intensity of the tail 
can be of the same magnitude as the intensity of the primary pulse. 

For the model under discussion it is also easy to find the ratio of
the energy ${\cal E}$ transferred by the tail term (beginning from the
time $t_1 + \Delta_2 (x)$) to the energy of the direct term ${\cal E}_0$,
namely
\[
 {{\cal E} \over {\cal E}_0} = \Bigl({2 M \over \tilde \Delta_2 (x)} 
 \Bigr)^2 F\Bigl({\tilde \Delta_2 (x) \over T} \Bigr), 
\]
where
\[
 F(\rho) = \rho + {\rho^2 \over \rho+1} - 2 \rho^2 \ln {\rho +1 \over 
 \rho}. 
\]
It is interesting to note that the function $F(\rho)$ has a maximum
at $\rho \approx 0.638$. 

\begin{figure}
\centerline{
\psfig{file=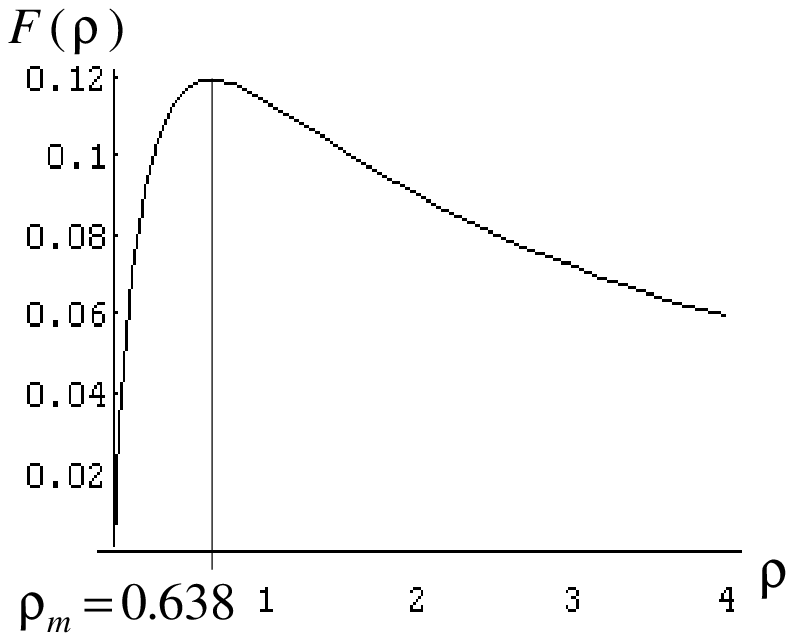,width=85mm} }
\caption{
The function $F(\rho)$ vs its argument $\rho$.
}
\end{figure}

Therefore, for the given model the amount of
energy transferred by the tail after the time $t_1 + \Delta_2 (x)$ is
maximal if the duration of the direct pulse is
$T \approx 1.57 \tilde \Delta_2 (x)$, and can in this case make up
nearly $10 \%$ of the energy of the direct pulse: 
${\cal E} \approx 0.119 {\cal E}_0 (2 M)^2 /\tilde \Delta_2^2 (x)$. 

As the quantity ${\cal F}$ does not explicitly depend on the mass of
the gravitational source, then its value for any particular case
(as 0.119 for the present example) is primarily determined by the
profile of the direct pulse and the spatial configuration of the
system consisting of the wave source, gravitational source and the
observer. Somewhat unexpected is the outcome that the magnitude of
the factor $(2M /\tilde \Delta_2)^2$, which characterizes the
influence of the gravitational source, can be of the order $1$,
even if the potential of the gravitational source is low everywhere,
i.e. $2 M \ll r_s$. This conclusion can be understood on the basis
of Ref. \cite{thorne} where it is shown that the tail field is
predominantly generated by the direct field in those regions where
gravitational focusing has deformed the geometry of the direct wave
fronts (see the shadowed region in Fig. 4). The deformation is
characterized by the focusing function $\alpha (\bar z,y)$. If the
points $\bar z$ and $y$ lie on a geodesic line which does not cross
the gravitational source, then $\alpha(\bar z,y)=0$. If the distance
$r_0$ of the wave source from the gravitational source is much
larger than the extension $r_s$ of the latter, then for the rays
originating in the wave source and passing through the gravitational
source it is valid $\alpha \sim Mr_0/r_s^2$. In the case of our
example the condition $2M /\tilde\Delta_2 \approx 1$ corresponds to
$\alpha \sim 1$. Let us note that the last effect is not revealed by
the traditional methods based on the expansion in terms of spherical
functions, as in this case it is natural to choose the effective
location of the wave source within the source of gravitation. Thus,
$\tilde \Delta_2 \approx r_s$ and $\alpha \sim M/r_s$, that is, the
intensity of the tail is proportional to the square of the potential
of the gravitational field on the surface of the source of
gravitation.

Similar estimates for the energy of scalar and electomagnetic dipole wave tails in a strong Schwarzschild field have been found respectively in Refs. \cite{malec1,malec2}.

\subsection{Constraints on applicability of the elaborated algorithm }

Relying on Ref. \cite{thorne}, we will now briefly analyse 
limitations to applicability of the above results. Firstly, note that the
conclusions of this paper were derived under the assumption that on the 
physically interesting spacetime domain the world function $\sigma$ is
single-valued, i.e. the geodesics originating from the point 
$y$ do not cross. If this condition is not fulfilled, then the 
developed algorithms are not
directly applicable. However, by virtue of the superposition
principle, valid because of linearity of the wave equation, these 
algorithms can be applied even when the world function is a 
multiple-valued function. Then 
the correct classical Green's function is the sum over all
distinct elementary classical Green's functions corresponding to all
distinct geodesics between the points $x$ and $y$ (see also \cite{dewitt}).

Crossing of geodesics would be caused by gravitational focusing; and
at the crossing point the exact factor $k(x,y)$ in ${\bf U}$ (see
(\cite{dewitt}))  would diverge. Thus, the criterion for no crossing 
is that $k(x,y)$ be finite along $C^+ (y)$. Let us now consider our 
first-order expression  (\ref{29}). 
This expression for $k(x,y)$ can
never diverge if the gravitational source is bounded. However, if the 
focusing function $\alpha (x,y) \equiv k(x,y)-1$ approaches 
unity, then the second- and 
higher-order effects will come into play, producing a divergence.
Thus, the constraint $\alpha (x,y) <1$ for all $x$ and $y$ is 
necessary, on the one hand, for the first-order analysis to be valid; 
one the other hand, it simultaneously avoids crossing of geodesics. 

For a wave-source in a circular orbit (the observer lying on the
plane of the orbit), Eq. (\ref{29}) gives 
$\text{max}~\alpha \sim (2M/r_s)(2r_0/r_s)$, 
where $M$ is the mass of the source of gravitation, $r_s$ is its linear
size and $r_0$ is the radius of the orbit. 
In this case the constraint $\text{max}~\alpha<1$ is significant: it
says that in order to avoid too much ray focusing, the wave-source 
and the gravitational source must not be too far apart
\be 
 {2M \over r_s} < {r_s \over 2 r_0}. \label{54}
\ee

Applying Eqs. (\ref{46}), (\ref{47}) and (\ref{49}) to a particular
model system, one should not overlook the fact that, although ${\bf E}$ 
is a first-order small quantity and therefore, when calculating the
integrals involved the spacetime may be considered as a flat background;
generally the retarded time $\tau (x)$ must be regarded as the retarded 
time on the curved spacetime because the integrands may be sensitive 
to the wave phase.

The quantities $\tilde \Delta_2 (x)$ and $\tilde \Delta_1 (x):=u(t_0 + 
\Delta_1 (x)) - u(t_0)$ offering their own inherent physical interest must 
generally also be calculated at least up to the accuracy of the 
first-order corrections. 

Let us now consider in more detail the time intervals
$\tilde \Delta_2 (x)$ and $\tilde \Delta_1 (x)$ for the binary system
described at the end of Sec. IV C.
We denote the origin of the coordinates at the center of the source of
gravitation by $z$ and consider only the case in which the points
$x$, $y$ and $z$ are aligned on a straight line. Evidently there are two
possible cases: 
(i) the point $z$ is placed between the points $x$ and $y$, (ii) the
point $y$ is placed between the points $x$ and $z$. Equations (\ref{36}) 
and (\ref{37}) yield the following conclusions.
In the case (i) the time intervals $\tilde \Delta_1 (x)$ and
$\tilde \Delta_2 (x)$ have maximal values:
\[
 \text{max}~\tilde \Delta_1 (x) = 2(r_0 - r_s) + \chi_1 (x),
\]
\[
 \text{max}~\tilde \Delta_2 (x) = 2(r_0 + r_s) + \chi_2 (x). 
\]
In the case (ii) the time intervals $\tilde \Delta_1 (x)$ and
$\tilde \Delta_2 (x)$ have minimal values:
\[
 \text{min}~\tilde \Delta_1 (x) =0, \quad \text{min}~\tilde \Delta_2 
 (x) = {r_s^2 \over 2 r_0} + \chi_3 (x). 
\]
Here the quantities $\chi_{\alpha} \sim 2M$, $\alpha=1,2,3$, denote the
corrections caused by the Shapiro effect within the inner region of the
binary system which are of the same order of magnitude as the
Shapiro time delay between the points $y$ and $z$. 

It is obvious that in Eqs. (\ref{46}), (\ref{47}) and (\ref{49}) the 
corrections caused by the inner Shapiro effect may be neglected only
in the low-frequency limit
\be
 \omega \ll {1 \over 2M}. \label{55}
\ee
The necessary condition (\ref{55}) is also sufficient, if the
inequality (\ref{54}) is simultaneously satisfied.

\section{SUMMARY AND DISCUSSION}

(i) On the basis of our earlier works \cite{laas,mankin3,mankin4}, 
we developed simple recurrent formulas, i.e. Eq. (\ref{12}), for 
calculating exact multipole solutions, i.e. Eq. (\ref{24}), of the 
electromagnetic wave equation on the background of a curved
spacetime which proceed from the classical Green's function
(i.e. the corresponding Hadamard coefficients).

(ii) The next main new contribution of this paper is formula 
(\ref{33}) with (\ref{30}) for the first-order tail term of 
the electromagnetic multipole wave in case of an arbitrary weak background
gravitational field. From these formulas will follow a number of physically
interesting effects.

Firstly, one can infer from Eqs. (\ref{30}) the earlier known 
(see \cite{gynther2,mankin6}) necessary and sufficient condition for
the validity of the Huygens' principle (in the sense of Hadamard) 
in the first approximation, namely $R_{ab}=0$.

Secondly, if the gravitational field source is spatially isolated,
then beginning from a certain instant of time the structure of the tail
accompanied by a pulse of electromagnetic radiation is completely
determined by the four-momentum of the source of gravitation, and
the higher multipole moments, including the angular momentum, 
do not at all influence the
structure of the tail in the first approximation.
For the scalar wave equation and for the
Green's function of the electromagnetic wave equation
essentially similar conclusions have been presented in Refs.  
\cite{mankin7} and \cite{mankin6}.

Thirdly, in Sec. IV B we discussed a further consequence of Eq.
(\ref{33}), namely, the time delay effect of the tail with respect 
to the direct pulse.
In the case of compact binary astrophysical objects the time interval
(blackout) in between the direct pulse and the tail caused by delaying
of the tail may be observable and can with the relative intensity
of the tail give, independently of the other methods, essential 
information about the characteristics of the physical system 
(the distance between compact objects, the mass of the source of 
gravitation, the orientation of the plane of the
orbit with respect to the observer, etc.). 
Earlier on the basis of the structure of the Green's function of
the wave equation in a weak Schwarschild field the time delay effect of
the tail and its possible astrophysical implications have been indicated in
the papers \cite{roe} and \cite{peters1}. Let us mention that some 
unpublished calculations by the present authors demonstrate the occurrence
of a similar delay effect also for linearized gravitational waves in the
second approximation (cf. \cite{peters2}). 
By the common approach in which the wave equation is solved by
separating the angular variables the delay effect, as a rule, remains 
unrevealed. This is due to the circumstance that ensuing from the 
symmetry of gravitational field (Schwarzschild, Kerr etc.), the origin 
of the spherical coordinates
whose worldline is simultaneously the worldline of the multipole
radiation source lies inside the source of gravitation where the
Ricci tensor does not vanish, $R_{ab} \not= 0$.

(iii) It is valid in the first approximation that the nonlocal
radiative wave-propagation correction at infinity takes a universal form
(\ref{45}) which is independent of the multipole order. A 
corresponding result in a weak 
Schwarzschild field in a slow-motion approximation has been
published in \cite{leonard}. Our Eq. (\ref{45}) generalizes 
their result in two aspects.
First, Eq. (\ref{45}) is valid in case of an arbitrary weak
gravitational field in the corresponding asymptotically flat spacetime. 
Second, there are no formal restrictions to the spatially 
bounded motion of the wave source,
e.g. the motion of the wave source can be relativistic.

(iv)  The intensity of the electromagnetic wave pulse emitted  by
the wave source within a compact binary in the vicinity of the geometric
shadow of the source of gravitation can be of the same magnitude 
as the intensity of the direct pulse, and the energy carried 
away by the tail can amount to $10 \%$ of
the energy of the low-frequency modes of the direct pulse. 

\acknowledgments

The present work was partially supported by the Estonian Science
Foundation grant No 4042 for which the authors extend their
gratitude. We would also like to thank Dr. E. Malec for a useful
discussion and express our regret that there was no reference to
the work \cite{malec1} in our paper \cite{mankin9}.

\subsection*{APPENDIX: NOTATION AND BASIC FORMULAS}

1. We adopt the metric signature $(+, -, -, -)$ 
with the Latin indices running and summing from $0$ to $3$. To 
abbreviate the notation of repeated tensor indices, we introduce 
multi-indices, e.g. $A(\mu):=(a_1 \dots a_\mu)$, $I(\mu):=(i_1 \dots i_\mu)$, 
$P(\mu):=(p_1 \dots p_\mu)$, etc. We apply extensively the technique of
two-point 
tensors (or bitensors) and follow closely the standard notation and 
index conventions \cite{dewitt}. To distinguish the tensor indices 
which refer to the point $x$ from those referring to the 
points $y$ and $z$ we use indices $a,b,\dots A(\mu),
B(\mu), \dots$ at $x$, indices  $i,j,\dots, I(\mu), J(\mu), \dots$ at $y$ 
and $p,q, \dots , P(\mu), Q(\mu), \dots$ at $z$. Thus $F^a_{I(3)}$ is a 
contravariant vector at $x$ and third-order covariant tensor at $y$. 
The index convention is also used for 
ordinary tensor fields, in which case it distinguishes the value of 
the component $v^a$ of a vector field ${\bf v}$ at $x$ from its 
value $v^i$ at $y$, and $G_{I(\mu)}$ denotes the covariant components 
of the tensor field ${\bf G}$ of rank $\mu$ at the point $y$. 

2. Symmetrization and antisymmetrization of the tensor indices is
denoted respectively by the parentheses and brackets. For example
$t^{(ab)}:=(t^{ab}+t^{ba})/2$ and $t^{[ab]}:=(t^{ab}-t^{ba})/2$.

3. Ordinary differentiation is denoted by a comma (,).
Covariant differentiation with respect to the Levi-Civita (metric) 
connection is denoted by $\nabla$ and semicolons, e.g. 
$\nabla_{x^a} f(x,y)=f_{;a} (x,y)$, $g^{ab} \nabla_{x^b} f(x,y) =
f^{;a} (x,y)\;$ and $\nabla_{I(2)} f(x,y)= \nabla_{y^{i_2}} 
\nabla_{y^{i_1}} f(x,y) = f_{;i_1 i_2} (x,y)$. Absolute
differentiation along a line $z(\la)$ is denoted by an overdot, i.e.
$\dot w^i=w^i_{;j} \dot z^j$. 

4. We use a system of units in which the speed of light and 
the gravitational constant are equal to $1$. 

5. The class of $C^{\infty}$ tensor 
fields of rank $m$ is denoted by ${\cal E}^m (\Omega)$ and the
subspace of ${\cal E}^m (\Omega)$ consisting of fields 
with compact support by ${\cal D}^m (\Omega)$. A tensor-valued
distribution ${\bf T} (x,y) \in {\cal D}'^m (\Omega)$ of rank $m$ at 
$x \in \Omega$ and of rank $k$ at $y \in \Omega$ is a continuous 
linear map ${\cal D}^m (\Omega) 
\to {\cal E}^k (\Omega)$. If $(\omega, \pi)$ is a coordinate chart
such that $x,y \in \omega$ and ${\bf \Phi} \in {\cal D}^m (\omega)$, 
then each  component of $(T_{I(k)}^{A(m)} (x,y), \phi_{A(m)} (x))$ 
is a (scalar-valued) 
tensor distribution (for a detailed discussion see \cite{fried}).
A set $\Omega_0 \subset \Omega$ is called past-compact if $J^- (x)
\cap \Omega_0$ is compact (or empty) for all $x \in \Omega$. We denote 
the class of distributions in ${\cal D}' (\Omega)$ with past-compact 
supports by ${\cal D}'^+ (\Omega)$. 

6. The world function biscalar $\sigma (x,y) \in C^{\infty} (\Omega
\times \Omega)$ is defined by the equations
\bea
 g^{ab} \nabla_a \sigma (x,y) \nabla_b \sigma (x,y) = 4 \sigma
(x,y),\label{2}\\
 \sigma(x,y)=\sigma(y,x), \qquad \lim_{x \to y} \sigma = 0.
 \eea
The world function $\sigma (x,y)$
is numerically equal to the square of the geodesic arc length
between the points $x$ and $y$, and is positive for timelike
intervals and negative for spacelike intervals.

7. The propagator of geodetic parallel displacement (also called the
transport
bitensor) is defined as a bitensor field ${\bf g} (x,y)$ of rank $m$ 
at both $x$ and $y$, which satisfies, in local coordinates, the
following differential equations and initial conditions: 
\bea 
 \sigma^{;a} \nabla_a g^{I(m)}_{B(m)} (x,y)=0,\label{4}\\
g^{I(m)}_{A(m)}(x,x)=\de^{i_1}_{a_1} \dots \de^{i_m}_{a_m}.
\eea

8. The surface distributions $\de_{\pm}^{(\mu)} 
(\si (x,y)), \mu=0,1,2, \dots $  are  defined by
\be 
 (\de_{\pm}^{(\mu)} (\si (x,y)), \phi (x)):= \lim_{\vaps \to +0}
\left(
 -{\partial \over \partial \vaps} \right)^{\mu}
\int_{C^{\pm}_{\vaps} (y)}
 \phi (x) \mu_{\si} (x) \label{5}
\ee
for all $\phi \in C_0^{\infty} (\Omega)$, if $\mu =0$, and for all
$\phi \in 
C_0^{\infty} (\Omega \backslash \{y \})$, if $\mu \geq 1$, where
$\mu_{\si} 
(x)$ is a Leray form for which 
\be 
 d \si (x,y) \wedge \mu_{\si} (x) = \mu(x), \label{6}
\ee
$\mu (x)$ being an invariant volume element and $C^{\pm}_{\vaps} (y)
= \{
x|x \in D^{\pm} (y), \si (x,y)=\vaps , \vaps>0\}$.  In
general, 
the integral (\ref{5}) can be evaluated by means of a partition of
unity subordinated to a covering of $\Omega$ by open coordinate
neighbourhoods in each of which
\be
 \mu (x) = \sqrt{g(x)} dx^0  \wedge dx^1 \wedge dx^2 \wedge dx^3,
\label{mux}
\ee
where $g(x):=|det(g_{ab} (x))|$ is the determinant of the metric
tensor. The relevant properties of the surface distributions 
are given in \cite{fried,gelfand}.

9. The $2$-form $\omega$ on the $2$-surface $S(y):= C^+ (y) \cap
C^- (x)$ in the integral (\ref{10-1}) is defined by 
\be
d \si (z,x) \wedge d \si (z,y) \wedge \omega (z) = \mu (z),
\label{6-1}
\ee
where $\mu(z)$ is an invariant volume element on $M$.

10. The Heaviside distributions are defined by
\be 
 (\Theta_{\pm} (\si (x,y)),\phi (x)):= \int_{J^{\pm} (y)} \phi (x)
\mu 
 (x), \label{7}
\ee
where $J^{\pm} (y) := D^{\pm} (y) \cup C^{\pm} (y)$. The relevant 
properties of the Heaviside distributions can be found in Ref.
\cite{fried,gelfand}.

11. The scalarized van Vleck determinant is a biscalar which is defined by
the equation
\be
k(x,y):={{|\det \si_{;ai} (x,y)|^{1/2}} \over
4|g(x)g(y)|^{1/4}},\label{8-3}
\ee

12. A {\it $\mu$th-order Green's function of the tensor wave operator
${\bf L}$ with respect to $(y, \Omega)$} is a tensor-valued
distribution ${\bf G} (x,y) \in {\cal D}'^1 (\Omega)$ of rank $1$
at $x$ and of rank $\mu +1$ at $y$ which satisfies the equation
\be
 LG^a_{iJ(\mu)} (x,y) = (-1)^{\mu} \nabla_{B(\mu)}\! \left[g^a_i (x,y) 
 g^{B(\mu)}_{J(\mu)} (x,y) \delta (x,y)\right], \label{tengrfun}
\ee
where $g^{B(\mu)}_{J(\mu)}$ is the tensor propagator of geodetic
parallel displacement.

\end{document}